\documentclass[11pt]{article}
\usepackage{amsmath,amsfonts,amsthm}
\usepackage{graphics, color}
\usepackage{natbib}

\RequirePackage[OT1]{fontenc}
\RequirePackage{amsthm,amsmath,natbib}
\usepackage{epsfig}
\usepackage{verbatim}
\usepackage{graphicx}
\usepackage{natbib}
\usepackage{amsmath,amsfonts,amsthm}
\usepackage{graphics, color}

\newcommand{\ex}[1]{\ensuremath{\mathbb{E}[#1]}}
\newcommand{\var}[1]{\ensuremath{\mathrm{Var}[#1]}}

\newcommand{\bd}[1]{\ensuremath{\mbox{\boldmath $#1$}}}

\begin{document}
%\begin{frontmatter}
\title{Estimating exposure response functions using ambient pollution concentrations}%\protect\thanksref{T1}}
%\runtitle{Estimating exposure response functions}
%\thankstext{T1}{Footnote to the title with the `thankstext' command.}

\author{Gavin Shaddick (1), Duncan Lee (2), James V. Zidek (3), \\Ruth Salway (1).\\
(1) University of Bath, (2) University of Glasgow,\\ (3) University of British Columbia.}

\maketitle

%\begin{aug}
%\author{\fnms{Gavin} \snm{Shaddick$^{1}$}\thanksref{t1}\ead[label=e1]{}}
%\author{\fnms{Duncan} \snm{Lee$^{2}$}}
%\author{\fnms{James} \snm{V. Zidek$^{3}$}}
%\and\author{\fnms{Ruth} \snm{Salway$^{1}$} },

%\affiliation{University of Bath$^{1}$ and University of Glasgow$^{2}$  and University of British Columbia$^{3}$}

%\address{Claverton Down\\ Bath\\ BA1 4EJ\\printead{e1}}

%\address{15 University Gardens\\ Glasgow\\ G12 8QQ\\printead{e2}}
%\affiliation{University of Glasgow}

%\address{333-6356 Agricultural Road\\Vancouver\\ BC\\V6T 1Z2\\printead{e3}}
%\affiliation{University of British Columbia}

%\address{Claverton Down\\ Bath\\ BA1 4EJ\\printead{e1}}
%\affiliation{University of Bath}

%\thankstext{t1}{Email address for corresponding author - g.shaddick@bath.ac.uk}
%\thankstext{t2}{First supporter of the project}
%\thankstext{t3}{Second supporter of the project}
%\runauthor{G. Shaddick et al.}

%\end{aug}

\begin{abstract}
This paper presents an approach to estimating the health effects of an environmental hazard. The approach is general in nature, but is applied here to the case of air pollution.  It uses a computer model involving ambient pollution and temperature inputs, to simulate the exposures experienced by individuals in an urban area, whilst incorporating the mechanisms that determine exposures.  The output from the model comprises a set of daily exposures for a sample of individuals from the population of interest. These daily exposures are approximated by  parametric distributions, so that the predictive exposure distribution of a randomly selected individual can be generated. These distributions are then incorporated into a hierarchical Bayesian framework (with inference using Markov Chain Monte Carlo simulation) in order to examine the relationship  between short-term changes in exposures and health outcomes, whilst making allowance for  long-term trends, seasonality, the effect of potential confounders and the possibility of ecological bias.\\

The paper applies this approach to particulate pollution (PM$_{10}$) and respiratory mortality counts for seniors in greater London ($\geq$65 years) during 1997. Within this substantive epidemiological study, the effects on health of ambient concentrations and (estimated) personal exposures are compared. The proposed model incorporates within day (or between individual) variability in personal exposures, which is compared to the more traditional approach of assuming a single pollution level applies to the entire population for each day. The relative risk associated with  (lag two) ambient concentrations of PM$_{10}$ was RR=1.02 (1.01-1.04). Individual exposures to PM$_{10}$ for this group (seniors) were  lower than the measured ambient concentrations, with the associated risk (RR=1.05 (1.01-1.09)) being higher than would be suggested by the traditional approach using ambient concentrations.

\end{abstract}
%\end{frontmatter}


\section{Introduction}

This paper addresses the differences between estimated associations observed in air pollution and human health studies.  The nature and magnitude of such associations will depend fundamentally on the nature of the study. Concentration response functions (CRFs)  are estimated primarily through epidemiological studies, that measure the association between ambient concentrations of pollution and a specified health outcome such as mortality (see \cite{daniels2004} for example). In contrast exposure response functions (ERFs) have been estimated through exposure chamber studies, where the physiological reactions of healthy subjects are assessed at safe levels of the pollutant (see \cite{ozone2006} for example). However ERFs cannot be ethically established in this way for those that matter most, namely  susceptible populations such as the very old and very young who are thought to be most adversely effected by pollution exposure. This paper presents a method for estimating the ERF based on ambient concentration measures.\\

We specifically consider the case of  particulate air pollution, which has attained great importance in both the health and regulatory contexts. For example they are listed in the USA as one of the so-called criteria pollutants that must be periodically reviewed. In fact such a review by the US Environmental Protection Agency led to a 2006 revision of the US air quality standards (\cite{PM04}), which require that in US urban areas daily ambient concentrations of PM$_{10}$ (particles no larger than $10$ microns in diameter) do not exceed $150$ $\mu$g/m$^3$ `more than once a year on average over three years. Human health concerns primarily underly these standards, as the US Clean Air Act of 1970 states they must be set and periodically reviewed to protect human health without consideration of cost while allowing for a margin of error. To a large extent then, regulatory policy is driven by health concerns.\\

In this paper we develop a model that estimates the ERF by relating personal exposures to daily health counts (aggregated over the entire population), and follows on from work by \cite{holloman2004} and \cite{shaddick05}. In particular we investigate the potential of using the pCNEM exposure simulator (\cite{zidek2005}) to generate personal exposures,  and compare the results to the CRFs estimated using routinely collected ambient levels. A case study is presented, in which we examine the relationships between (daily) respiratory mortality and both ambient levels (CRF) and individual (simulated) exposures (ERF) of particulate matter (PM$_{10}$) for  seniors ($\geq$ 65 years) in Greater London (for 1997). The remainder of the paper is organised as follows. Section 2 provides the background and motivation for this work, while section 3 describes the proposed  model and section 4 presents the case study of data from Greater London. Throughout we adopt a Bayesian approach to modelling, with inference  using Markov Chain Monte Carlo simulation.  Section 5 provides a concluding discussion.

\section{Background}
The majority of  studies relating air pollution with detrimental effects on  health have focused on the short-term relationship, and relate to ecological data from a geographical region $\mathcal{R}$ (such as a city) for $n$ consecutive days. Such relationships are typically estimated using non-independent regression or `time series' models, that regress daily mortality counts $\mathbf{y}=(y_{1},\ldots,y_{n})_{n\times1}$ against air pollution levels and a vector of $q$ covariates. The latter  are denoted by $Z=(\mathbf{z}_{1}^{\scriptsize\mbox{T}\normalsize},\ldots,\mathbf{z}_{n}^{\scriptsize\mbox{T}\normalsize}) ^{\scriptsize\mbox{T}\normalsize}_{n\times q}$,  where $\mathbf{z}_{t}^{\scriptsize\mbox{T}\normalsize}=(z_{t1},\ldots,z_{tq})$ represent the realisations for day $t$. These covariates include meteorological conditions such as temperature together with smooth functions of calendar time which model unmeasured risk factors that induce long-term trends, seasonal variation, over-dispersion and temporal correlation into the mortality data. In general only ambient pollution levels, $x^A_{jt}$, measured by a network of  $k$ fixed site monitors located across the study region are available. A daily average $x^A_{t}=(1/k)\sum_{j=1}^{k}x^A_{jt}$ is typically calculated across these $k$ spatial observations, which is related to the mortality counts using Poisson linear or additive models. A Bayesian implementation of the former is given by

\begin{eqnarray}
y_{t}&\sim&\mbox{Poisson}(\mu_{t})~~~~\mbox{for}~~t=1,\ldots,n,\nonumber\\
\ln(\mu_{t})&=&x^A_{t-l}\gamma+\mathbf{z}_{t}^{\scriptsize\mbox{T}\normalsize}\bd{\alpha},
\label{equation standard model}\\
\bd{\beta}=(\gamma,\bd{\alpha})&\sim&\mbox{N}(\bd{\mu}_{\beta},\Sigma_{\beta}),\nonumber
\end{eqnarray}

\noindent where the Gaussian prior for $\bd{\beta}$ is typically vague. In this model the association between ambient pollution levels (at lag $l$) and mortality is represented by $\gamma$, and is of interest for regulatory purposes primarily because it is only ambient pollution levels that are routinely measured.
However to obtain more conclusive evidence of the public health impact of air pollution, exposures actually experienced by individuals are required, enabling the ERF to be estimated. Ideally, this would be done within individual level studies conducted under strict conditions, such as in randomised controlled trials. However due to the difficulties and costs of obtaining personal pollution exposures and individual health events with histories detailed enough to deal with confounding,  even observational studies conducted at the individual level are rare (a few examples are given by \cite{neas1999}, \cite{yu2000} and \cite{hoek2002}).\\

Personal exposures are based on indoor as well as outdoor sources, and are likely to be different from ambient levels (see for example \cite{dockery1981} and \cite{lioy1990}) because the population spend a large proportion of their time indoors. However such exposures are prohibitively expensive to obtain for a large sample of the population, and consequently only a few studies (see for example \cite{lioy1990} and \cite{ozkaynak1996}) have collected individual level data which has been restricted to less than 50 daily exposures over up to 120 days. As a result few studies have estimated the association between personal exposures and mortality, with one of the first being that of \cite{dominici2000a} who analyse data from Baltimore. However they report that individual pollution exposures are not available for Baltimore, and instead estimate them by transporting a
linear relationship between ambient levels and average exposures from five external data sets. However due to the difficulties in obtaining
personal exposure data the samples of personal exposures are small, which may lead to problems when assuming that the sample represents the exposure of the entire population to which the health counts relate. \\

If a sample of actual personal exposures
is not available, it may be possible to use simulated exposures. Such exposures have been generated by models such as SHEDS-PM (\cite{burke2001}), APEX (\cite{apex}) and pCNEM (\cite{zidek2005}), and have played an important role in formulating air quality criteria resulting in two important applications. The first and most widely used is that they can evaluate abatement strategies (e.g. regulations and mandatory surveillance), by running the model before and after hypothetical changes in policy (see  \cite{zidek2007}). The Environmental Protection Agency in the US have used such models to estimate carbon monoxide and ozone exposures (pNEM, a fore-runner to pCNEM \cite{law1997}) at the population level, while particulate matter  has been modelled using SHEDS. In addition the latest ozone criterion document (\cite{ozone2006}) makes use of the APEX model, while \cite{zidek2007} used pCNEM to forecast personal exposures of PM$_{10}$ after a theoretical `roll-back' programme. Although they may differ in certain often fundamental respects, all of the simulators have important
conceptual elements in common. Namely, they estimate the cumulative exposure experienced by individuals as they pass through different (micro-) environments, for example a car, house, street, which is calculated from the different pollution levels in each of these environments. The second application,  which is proposed in this paper has attracted far less attention and uses exposure simulators to generate more accurate estimates of population exposures. \cite{holloman2004} relate simulated individual exposures to mortality data from North Carolina, the former of which are generated from a deterministic simplification of the SHEDS-PM simulator (\cite{burke2001}). In a forerunner to this work, \cite{shaddick05} used simulated daily exposures and related them to mortality counts in London, observing an increased relative risk
 compared with ambient concentrations, although this was accompanied by a widening of the 95\% credible interval.\\

Although the studies of \cite{dominici2000a} and \cite{holloman2004} have related individual exposures to ecological mortality counts, the models used  have a number of limitations. Primarily they summarise  daily exposure distributions by a simple average while not allowing for the possibility of ecological bias (\cite{wakefield2001}), which may arise when variation in the exposures is ignored. Comprehensive reviews of the relationship between aggregate and individual models as well as ecological bias are given by \cite{richardson1987}, \cite{zeger2000} and \cite{wakefield2001}. When extending this simple average to allow for exposure variability both papers make a Gaussian assumption, which is likely to  be inappropriate for non-negative environmental exposures of this type (see \cite{ott1990}).

\section{Statistical modelling}
 Here we propose a two stage modelling strategy for generating and relating personal exposures to mortality, that differs from the `all at once' approach adopted by \cite{holloman2004}. The first sub-section describes the pCNEM simulator, while the second proposes a model to relate these exposures to aggregated mortality counts.

\subsection{Stage 1: Estimating average population exposure}
The pCNEM simulator and a previous implementation using data from London
is described in \cite{zidek2005}. Described simply, it generates a
sequence of pollutant concentrations to which a randomly selected
individual is exposed over time. This sequence is termed the {\em
personal exposure sequence}. The generation is a fairly complex
stochastic process that follows the randomly selected individual in
their activities over the period of the simulation. The individual is
thought of as visiting one {\em microenvironment} after another
as activities change through time. The simulator  has two major
tasks; (i) to create estimates of the  levels of pollution in each
microenvironment over time and (ii)  to generate an activity sequence for a
randomly selected individual. The individual's cumulative level of
exposure is then calculated by tracking them through their different
activity levels within the microenvironments.

\subsection{Stage 2: Estimating the effects of exposure to health}

We propose a model that extends the standard approach of representing daily pollution levels by a single value (for example the mean). Instead we assign a probability distribution to the daily exposures. For clarity, the lag notation, $X_{t-l}$, is dropped for the remainder of this subsection. Assuming the standard log-linear model as in (\ref{equation standard model}), ecological bias can be modelled (\cite{salway2007}) by considering the mean function,

\begin{eqnarray}
\label{commonform}
\mu_{t}&=&\mathbb{E}_{X_t} [ \mathbf{z}_{t}^{\scriptsize\mbox{T}\normalsize}\bd{\alpha}+ {\exp(g(\gamma X_t)}] ,\\ \nonumber
 &=&\exp( \mathbf{z}_{t}^{\scriptsize\mbox{T}\normalsize}\bd{\alpha})  \mathbb{E}_{X_t} [\exp (g(\gamma X_t))],
\end{eqnarray}

\noindent where $X_t$ represents daily pollution levels which come from a distribution $f(x_{t}|\bd{\lambda})$. The exposure response function is represented by $g$, while $\gamma$ is the log of the relative risk associated with a unit increase in pollution, and $\mathbf{z}_{t}^{\scriptsize\mbox{T}\normalsize}\bd{\alpha}$ models the other covariates. \\

\noindent We first consider the common case of $g(x)=x$, which  gives

\begin{eqnarray}
\label{commonform1b}
\mu_{t}=\exp(\mathbf{z}_{t}^{\scriptsize\mbox{T}\normalsize}\bd{\alpha}) \mathbb{E}_{X_t}[\exp(\gamma X_t)].
\end{eqnarray}

\noindent In addition to this linear response function, the traditional approach also assumes that $X_t$ can be represented by a single value $\lambda_t$, (for example the daily mean), meaning that (\ref{commonform1b}) simplifies to

\begin{eqnarray}
\label{commonform2}
\mu_t =  \exp(\mathbf{z}_{t}^{\scriptsize\mbox{T}\normalsize}\bd{\alpha}) \exp(\gamma \lambda_t)
\end{eqnarray}

\noindent which does not incorporate exposure variability and  thus may be effected by ecological bias. \cite{richardson1987} and \cite{salway2007} model ecological bias parametrically in this context by incorporating higher order moments (for example the variance) of the exposure distribution $f(x_{t}|\bd{\lambda})$ in the linear predictor, in addition to the mean. An alternative approach that addresses ecological bias is the aggregate approach of \cite{prenticesheppard95}), however the lack of a tractable distribution for the disease counts when using this method means it would not fit naturally into the fully parametric Bayesian framework used here.\\

If $X_t$ is  normally distributed, $X_t \sim N(\lambda_{t}^{(1)}, \lambda_{t}^{(2)})$, then the effects of exposure variability and ecological bias can be modelled exactly by the mean function

\begin{eqnarray}
\mu_t = \exp(\mathbf{z}_{t}^{\scriptsize\mbox{T}\normalsize}\bd{\alpha}) \exp( \gamma \lambda_{t}^{(1)} + \gamma^2 \lambda_{t}^{(2)}/2).
\label{equation parametric normal}
\end{eqnarray}

\noindent  If the daily exposures $X_{t}$ do not follow a normal distribution equation (\ref{equation parametric normal})  will be a second order approximation to the true model, which  is likely to be adequate provided the distribution of $X_t$ is not heavily skewed. \cite{ott1990} has shown that a log-normal distribution is appropriate for modelling environmental exposures because in addition to the desirable properties of right-skew and non-negativity, there is justification in terms of the physical explanation of atmospheric chemistry.  However ecological bias cannot be modelled in this way if the exposures are assumed to follow a log-normal distribution, because the moment-generating function does not exist. \cite{salway2007} suggest that if $\gamma$ is small (such as in air pollution and mortality studies) a three term Taylor approximation 

\begin{eqnarray}
\mu_t \approx \exp( \mathbf{z}_{t}^{\scriptsize\mbox{T}\normalsize}\bd{\alpha})\exp(\gamma \lambda_{t}^{(1)} + \gamma^{2}\lambda_{t}^{(2)}/2+ \gamma^{3}\lambda_{t}^{(3)}/6),\label{ruth_model}
\end{eqnarray}

\noindent can be used to model ecological bias. Here $(\lambda_{t}^{(1)},\lambda_{t}^{(2)},\lambda_{t}^{(3)})$ represent the first three central moments of the log-normal exposure distribution (that is $\lambda_{t}^{(r)}=\ex{(x_{ti}-\lambda_{t}^{(1)})^{r}}$, meaning that $\lambda_{t}^{(1)}=\ex{x_{ti}}$,
$\lambda_{t}^{(2)}=\var{x_{ti}}$ and $\lambda_{t}^{(3)}=\lambda^{(2)}_{t}/ \lambda_{t}^{(1)}(\lambda^{(2)}_{t}/ \lambda_{t}^{(1)}+3)$.\\

\subsubsection{Form of the exposure response function (ERF)}

The common simplification that $g(x)=x$ is not appropriate for air pollution studies, because there must eventually be an upper bound on the effect that air pollution can have on health. A more sensible approach is to consider a general function $g$ that satisfies the desirable requirements of: (i) boundedness; (ii) increasing monotonicity; (iii) smoothness (thrice differentiability); and (iv) $g(0)=0$.  Note that these properties are not commonly enforced on CRFs estimated for ambient pollution levels using generalised additive models (see for example \cite{dominici2002b}). These assumptions allow $\exp(g(\gamma X_{t}))$ to be approximated using a three term Taylor expansion of the form

\begin{eqnarray} \nonumber
\mu_t&\doteq&\exp(\mathbf{z}_{t}^{\scriptsize\mbox{T}\normalsize}\bd{\alpha})\mathbb{E}_{X_t} [\exp\{g(\gamma X_t)\}],\\ \nonumber
&\approx&\exp(\mathbf{z}_{t}^{\scriptsize\mbox{T}\normalsize}\bd{\alpha})\exp(g(\gamma\lambda_t^{(1)}))(1+ \gamma^2 g^{(2)}(\gamma\lambda_t^{(1)})\lambda_t^{(2)} + \gamma^3 g^{(3)}(\gamma\lambda_t^{(1)}) \lambda_t^{(3)}), \\
&\approx&  \exp(\mathbf{z}_{t}^{\scriptsize\mbox{T}\normalsize}\bd{\alpha})
\exp(       g(\gamma \lambda_{t}^{(1)}) +\gamma^{2} g^{(2)}(\gamma\lambda_t^{(1)})\lambda_t^{(2)} + \gamma^{3} g^{(3)}(\gamma\lambda_t^{(1)}) \lambda_t^{(3)}         ),
\end{eqnarray}

\noindent where again $(\lambda_{t}^{(1)},\lambda_{t}^{(2)},\lambda_{t}^{(3)})$ represent the first three central moments of the log-normal exposure distribution.  Ideally the values of the parameters $g^{(r)}$ would be estimated within the MCMC simulation, although in practice  it is unlikely that there would be enough information to estimate them.  Our preliminary analysis suggests the first term $g(\gamma\lambda_{t}^{(1)})$ can be well approximated by $g(\gamma\lambda_t^{(1)}) = \gamma\lambda_t^{(1)}$, and the lack of information to accurately estimate the derivatives of $g$ leads us to use the values suggested by  \cite{salway2007}; $g^{(2)}(\gamma\lambda_t^{(1)}) =1/2$ and $g^{(3)}(\gamma\lambda_t^{(1)})=1/6$, who proposed such a structure to  allow for ecological bias as in equation (\ref{ruth_model}). Note that the effect of the latter two terms of this approximation is likely to be small given the expected (small) values of $\gamma$.

\subsubsection{Model relating daily distributions of individual exposures to aggregate level counts}
We model the daily exposure distributions as log-normal and adopt the ecological bias correction from (\ref{ruth_model}). Our work thus extends that of \cite{dominici2000a} and \cite{holloman2004} who assumed a normal distribution for the distribution of $X_t$, but used a mean function of the form of equation (\ref{commonform2}), with no allowance for the possibility of ecological bias. This results in the following model for relating aggregate mortality counts to a sample of personal pollution exposures.

\begin{eqnarray}
y_{t}&\sim&\mbox{Poisson}(\mu_{t})~~~~
\mbox{for}~~t=1,\ldots,n,\nonumber\\
\ln(\mu_{t})&=&\lambda_{t-l}^{(1)}\gamma+
\lambda^{(2)}_{t-l}\gamma^{2}/2+
\lambda_{t-l}^{(3)}\gamma^{3}/6
+\mathbf{z}_{t}^{\scriptsize\mbox{T}
\normalsize}\bd{\alpha},\nonumber\\
\bd{\beta}=(\gamma,\bd{\alpha})&\sim&
\mbox{N}(\bd{\mu}_{\beta},\Sigma_{\beta}),\nonumber\\
x_{it}&\sim&\mbox{Log-Normal}
(\lambda_{t}^{(1)},\lambda^{(2)}_{t})
~~~~\mbox{for}~~i=1,\ldots,k_{t},\label{equation parametric lognormal model}\\
\lambda_{t}^{(1)}&\sim&\mbox{N}(\xi, \sigma^{2}),\nonumber\\
\lambda_{t}^{(2)}&\sim&\mbox{N}(s^{2}, \tau^{2})_{I[\lambda_{t}^{(2)}>0]},\nonumber\\
\sigma^{2}&\sim&\mbox{Inverse-Gamma}(0.001, 0.001),\nonumber\\
\tau^{2}&\sim&\mbox{Inverse-Gamma}(0.001, 0.001),\nonumber
\end{eqnarray}

\noindent where $x_{it}$ denotes the exposure experienced by individual $i$ on day $t$. The value of the lag $l$ is typically chosen as one or two (\cite{dominici2000b}). Note that (\ref{equation parametric lognormal model}) allows for sample error in the first two central moments of the daily exposure distribution, which are treated as unknown and assigned Gaussian priors based on prior knowledge of the mean values. Theoretically different prior means could be assigned to each day (i.e. $\lambda_{t}^{(1)}\sim \mbox{N}(\xi_t, \sigma^{2})$), but as the information required to sensibly choose values for these is unlikely to be available we use a common underlying mean for all days. The exposure variance $\lambda_{t}^{(2)}$ is assigned a truncated Gaussian prior because its expected value can be directly specified as a parameter, which would not be the case for standard variance priors such as inverse-gamma.

\section{Case study}
In this section we apply a series of models to data from Greater London and
are motivated by three aims: (i) demonstrate the potential of the
pCNEM exposure simulator for generating individual exposures for use
in air pollution and mortality studies; (ii) investigate the
differences between the effects of ambient pollution and personal
exposures on mortality; (iii) compare the performance of the
log-normal  model (\ref{equation parametric
lognormal model}) against simpler alternatives that have previously been adopted. The first sub-section
describes the data used in this case study, the second presents the results of pCNEM and describes the  statistical models that were considered, while the
third discusses the Bayesian model building process. The final two sections investigate the effects of ambient levels and personal
exposures on public health.

\subsection{Description of the data}
The data used in this case study relate to daily observations from
the Greater London area during the period 2$^{\mbox{nd}}$ January
1997 until 30$^{\mbox{th}}$ December 1997. The health data comprise
daily counts of respiratory mortality for seniors ($\geq$65 years)
drawn from the population living within Greater London. The
pollution data relate to particulate matter measured as PM$_{10}$,
and the pCNEM exposure simulator uses ambient levels measured at
eight spatial locations throughout Greater London. When using the simpler models, average ambient pollution
levels are calculated as the mean level over the eight monitoring
sites. This spatial average has also been used by
\cite{katsouyanni1996} and \cite{dominici2000b}, and is likely to
introduce minimal exposure error because PM$_{10}$ levels in London
between 1994 and 1997 exhibit little spatial variation
(\cite{shaddick2002}). Meteorological data (measured at Heathrow
airport) are also available for Greater London, including indices
of temperature, rainfall, wind speed and sunshine.

\subsection{Models}
\begin{figure}
\caption{Panel (a) shows boxplots of the 800 personal exposures by day. For clarity, only every second day is shown and the `whisker' component is removed. Panel (b) shows  the relationship between mean ambient levels and mean daily exposure. In both, personal exposures and  ambient concentrations (of PM$_{10}$) are measured in $\mu$g/m$^3$. }

\scalebox{0.69}{\includegraphics{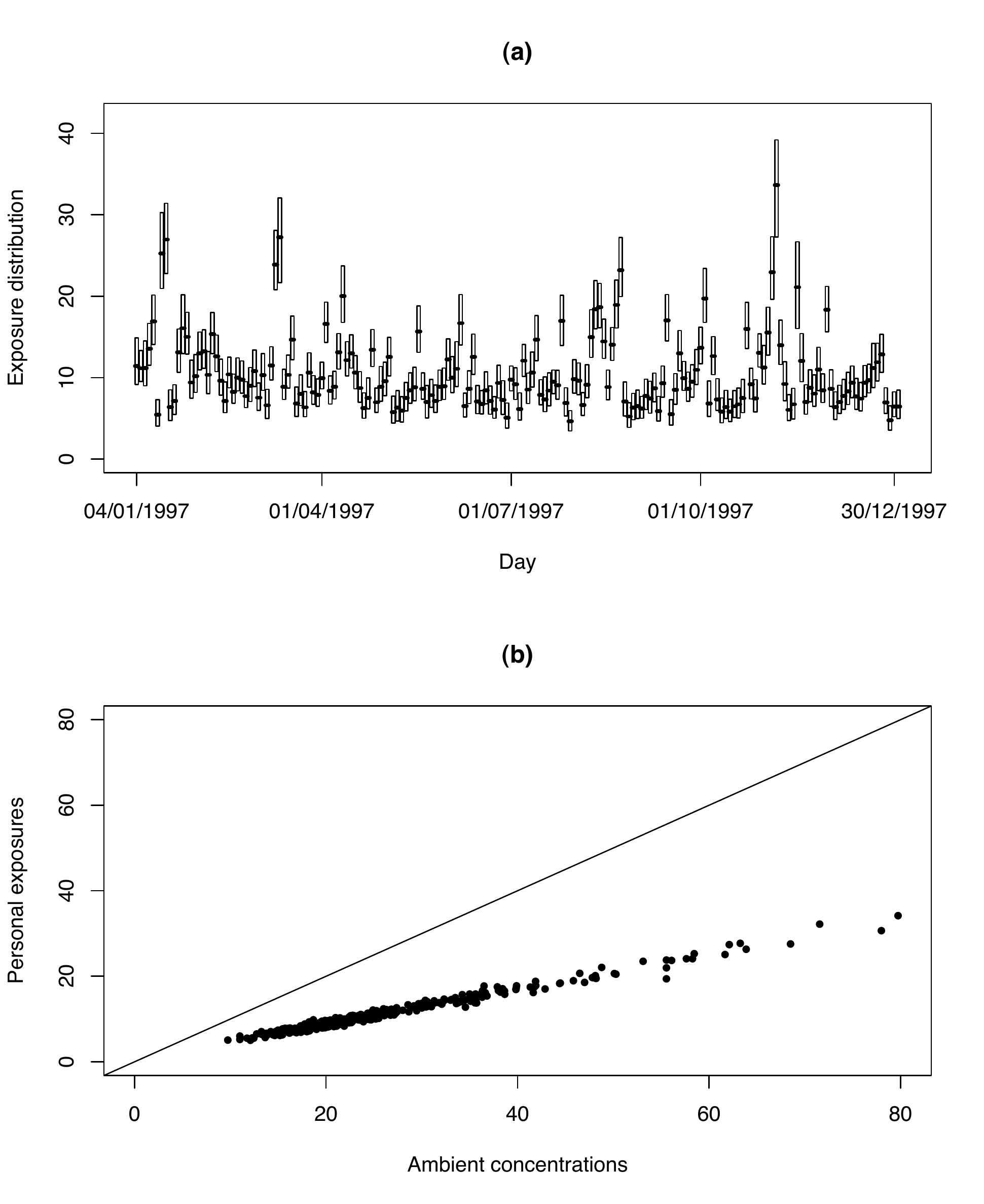}}
\label{graph_pcnem_data}
\end{figure}

The pCNEM simulator was run with ambient PM$_{10}$ levels from eight monitoring sites together with  maximum daily temperatures. Further  details of this implementation can be found in \cite{zidek2005}.  The  output consists of 100 replicates generated for each of the eight exposure districts (defined as areas around each of the monitoring sites), giving a total of 800 samples of exposures for each day. The distribution of daily simulated exposures are shown in panel (a) of Figure \ref{graph_pcnem_data}, while a comparison of ambient and personal exposures is presented in panel (b). These empirical exposure distributions are then modelled parametrically (normal or log-normal according to the case below) in our Bayesian hierarchical model. \\

In order to investigate the relationships between mortality and both personal exposure as well as ambient pollution levels, we apply a series of models to the data from Greater London.

\begin{itemize}
\item[(i)] The standard Poisson regression model (see equation \ref{equation standard model}) where the daily exposure is fixed at a single measure of ambient concentration.

\item[(ii)] The standard Poisson regression model (again the form of equation \ref{equation standard model}) where the daily exposure is fixed at the mean of the empirical distribution of daily exposures.

\item[(iii)] The normal exposure model (as in  \cite{holloman2004}) where daily exposures are assumed to follow a normal distribution.

\item[(iv)] The  log-normal exposure model (as in equation (\ref{equation parametric lognormal model})) where daily exposures follow a log-normal distribution.
\end{itemize}

In all models the mortality data are related to pollution levels lagged by two days. In order to investigate possible different effects related to indoor and outdoor sources of pollution exposure, we also run models (ii) to (iv) on these separate components of personal exposures.

\subsection{Modelling underlying temporal patterns and meteorological covariates}
The same set of covariates are used in each of these models, as only a single set of mortality data are used. These covariates are given by

$$\mathbf{z}_{t}^{\scriptsize\mbox{T}\normalsize}\bd{\alpha}=
\alpha_{1}+S(t|11,\bd{\alpha}_{2})+
S(\mbox{temperature}_{t}|2,\bd{\alpha}_{3})$$

\noindent where $S(var|df,\bd{\alpha}_{j})$ denotes a natural cubic spline of the variable \textit{var} with \textit{df} degrees of freedom. Natural cubic splines are used here in preference over non-parametric alternatives, because they are less cumbersome to implement within a Bayesian analysis using MCMC simulation. The covariates ($\mathbf{z}_{t}^{\scriptsize\mbox{T}\normalsize}\bd{\alpha}$) are included to model any trend, seasonal variation and temporal correlation present in the respiratory mortality series, and are chosen using a fully Bayesian model building process. The mortality data (not shown) exhibit a pronounced yearly cycle, with much less prominent cycles at periods of a half, quarter and eighth of a year. As the most prominent feature of the data is the yearly cycle, we began the model building process by accounting for it using daily mean temperature (which also has a pronounced yearly cycle). The temperature covariate was added to the linear predictor (which initially contained an overall intercept term) as either a  linear term or a smooth function, the latter of which was implemented  with a variable number of degrees of freedom. This process was repeated for a number of different lags and moving averages, and the fit to the data was compared using the deviance information criterion (DIC, \cite{spiegelhalter2002}) and examining plots of the standardised residuals (based on posterior medians). From this analysis a smooth function of the same day's temperature with two degrees of freedom was chosen, because it has the lowest DIC and visually appeared to remove the largest proportion of the yearly cycle from the residuals.  Meteorological indices of rainfall, wind speed and sunshine were also included in the model, but as they exhibited no relationship with mortality at any lag, they were not used in the final models. After including the temperature covariate the residuals still exhibited a (less) pronounced yearly cycle, together with cycles at higher frequencies. Such cyclical trends can be modelled by functions of calendar time, with a range of rigid or  free-form parametric and non-parametric alternatives being used in the literature. In this study we model this cyclical trend using a smooth function of calendar time implemented with natural cubic splines, because it is less rigid than parametric alternatives (such as pairs of sine and cosine terms) and has become the method of choice in most recent studies (see for example \cite{daniels2004}). In common with the temperature covariate we choose the degrees of freedom by calculating the DIC criterion and examining plots of the standardised residuals, which results in 11 degrees of freedom being selected. The adequacy of our chosen covariates is then assessed using the posterior predictive method of \cite{gelman1996}, which is a Bayesian model checking technique that predicts the posterior distribution of functions of the data and parameters. For example letting $r_{t}$ denote the standardised residual on day $t$, that is $r_{t}=(y_{t}-\mu_{t})/\sqrt{\mu_{t}}$, its posterior predictive distribution is given by

$$f(r_{t}|\mathbf{y},X)=
\int_{\bd{\theta}}f(r_{t}|\bd{\theta},\mathbf{y},X)f(\bd{\theta}|\mathbf{y},X)d\bd{\theta}$$

\noindent where $\bd{\theta}$ is generic notation for the model parameters and ($\mathbf{y},X$) denote the mortality and pollution data respectively. The adequacy of the covariates can be assessed by examining the posterior predictive distributions for these daily residuals (Figure 2(a)), as well as their autocorrelation sequence (Figure 2(b)). The residual distributions in Figure 2(a) show no clear pattern, suggesting that the covariates are likely to have adequately removed the trends and structure in the mortality data. The adequacy of the covariates is re-enforced by Figure 2(b), which shows that the standardised residuals exhibit little or no correlation. Now that the covariates have been chosen the effect of both ambient pollution levels and personal exposures at a number of lags and moving averages on mortality can be investigated. Lags of two days were chosen for both ambient levels and actual exposures because they have the lowest DIC for both pollution metrics, and exhibit the largest relationship with respiratory mortality. A sensitivity analysis revealed that the relationship between the effects of ambient levels and personal exposures are robust to the choice of lag. A sensitivity analysis also showed that the choice of $\epsilon$ in the inverse-gamma$(\epsilon, \epsilon)$ variance priors did not effect the posterior variances.\\

\begin{figure}
\caption{Posterior predictive distributions used for model checking. Panel (a) shows  the
standardised residuals. For clarity only every second day's
distribution is shown and the `whisker' component is omitted. Panel (b) shows  the
autocorrelation sequence. The numbers denote the median values.}
\scalebox{0.69}{\includegraphics{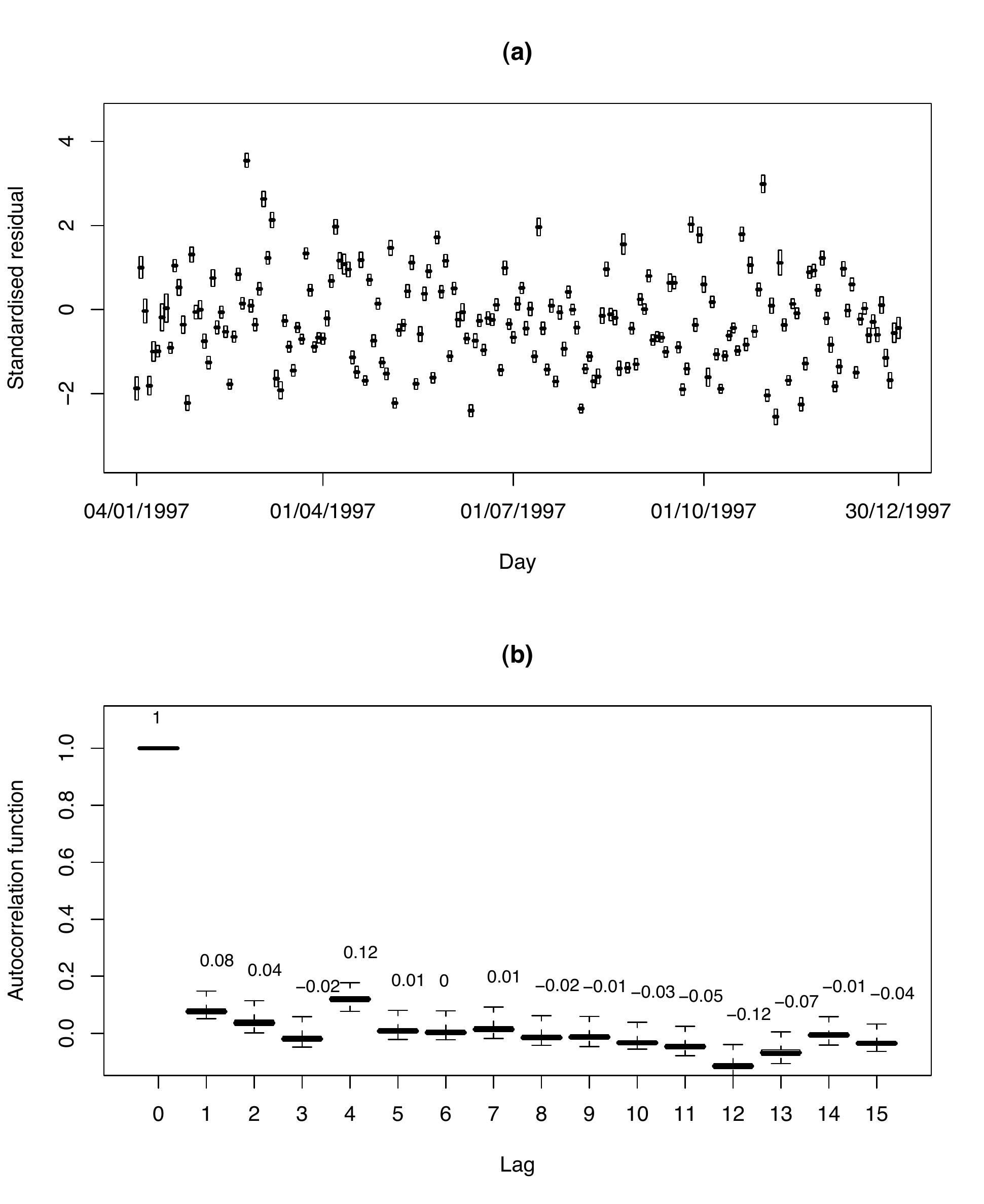}}
\label{graph_residual_distribution}
\end{figure}

Inference for all models is based on MCMC simulation, using a mixture of Gibbs sampling steps and block Metropolis-Hastings moves based on random walk proposals. In each case inference about the posterior distribution is based on 20,000 iterations from two Markov chains, initialised from dispersed locations in the sample space (in all cases the starting distributions are an overdispersed version of the prior). Both chains are burnt in for 20,000 iterations, by which point convergence was assessed to have been reached using the diagnostic methods of \cite{gelman2003}. After convergence is reached each chain is run for a further 250,000 iterations, which are thinned by 25 to all but remove the autocorrelation, resulting in 10,000 samples from each chain. To aid convergence of the Markov chains the covariates (the basis functions for the natural cubic splines of calendar time and temperature) are standardised to have a mean of zero and a standard deviation of one before inclusion in the model (and are subsequently back-transformed when obtaining results from the posterior distribution).

\subsection{Relationships between  pollution  and mortality}

\begin{figure}
\caption{Personal exposure distribution for 11$^{th}$
April 1997. Panel (a) depicts the empirical distribution, panel (b)
shows the effect of using a single value for daily exposure (model ii), panel (c) is the normal exposure model (model iii) and panel (d) is the log-normal exposure model (model iv).}
\scalebox{0.7}{\includegraphics{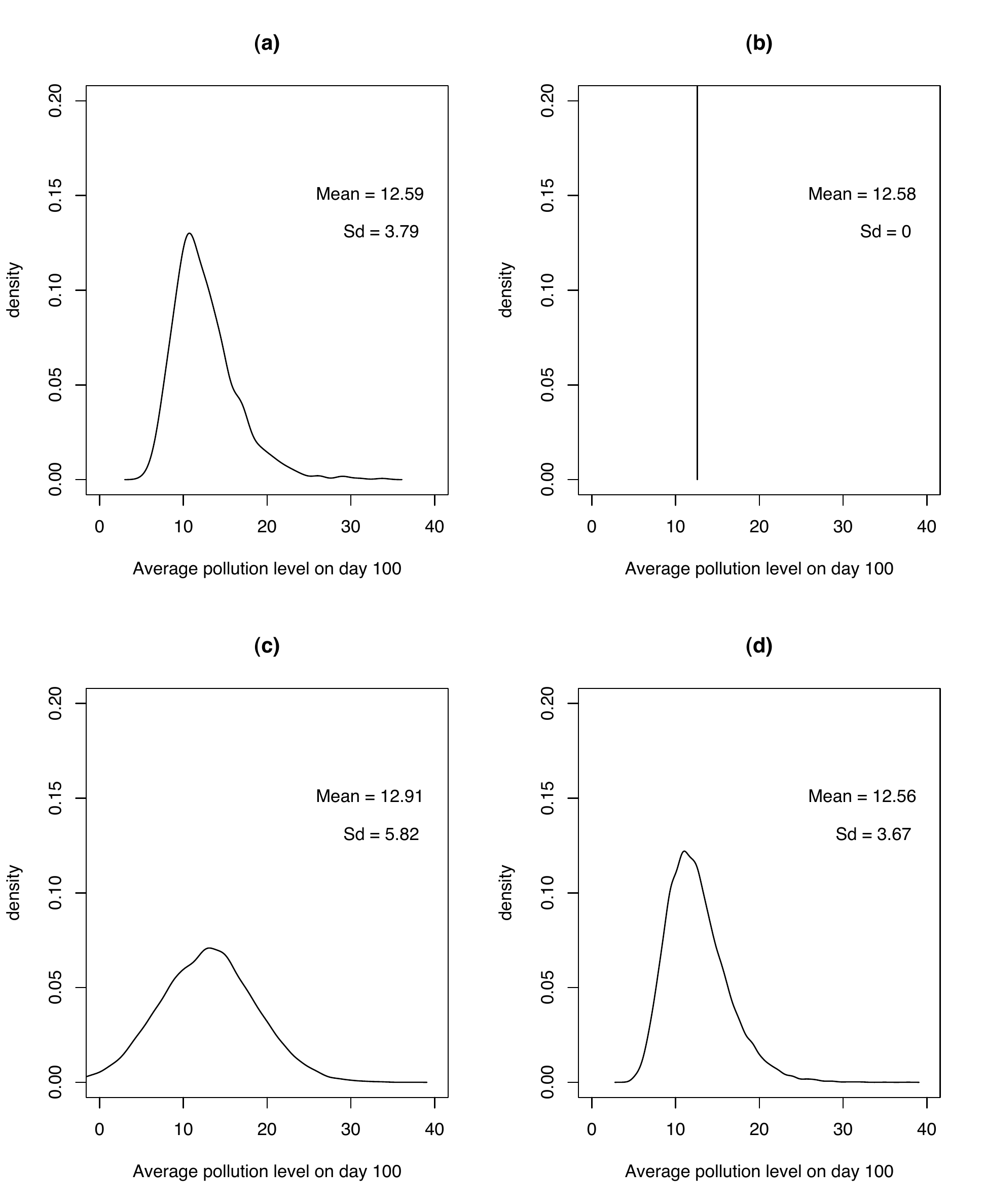}}
\label{graph_exposure_distribution}
\end{figure}

The estimated relationships between mortality and both personal exposure to  and ambient levels of  PM$_{10}$ are given in Table \ref{table_RR}, and are presented on the relative risk scale for an increase in 10 units of pollution. An increased risk  was observed in association with ambient concentrations,  RR=1.02  with a 95\% credible interval  of 1.01-1.04 . 
Relationships between personal exposures and aggregate level mortality counts are investigated using a standard Poisson regression model (ii),  the log-normal exposure model (iv) and the alternative based on a normal exposure distribution (iii). Table \ref{table_RR} shows that for personal exposures using both the log-normal and fixed exposure models, a 10 $\mu$g/m$^3$ increase in PM$_{10}$ gives a relative risk of 1.05 (1.01-1.09), which compares with a value of just over $1.04\%$ found by \cite{dominici2000a}. However in contrast, the  normal exposure model  estimates a relative risk of 1.07 (1.02-1.12), which is at odds with the other estimates. The normal exposure model (as well as the fixed exposure model) also does not capture the shape of the exposure distribution across the population, a point which is illustrated in Figure \ref{graph_exposure_distribution}. This shows the empirical distribution of the personal exposures for a randomly selected day (11$^{th}$ April 1997), and compares that to the estimated posterior predictive distributions from the proposed models. The empirical distribution is shown in panel (a), an illustration of using a single value rather than a distribution is shown in panel (b), while panels (c) and (d)  contain the posterior predictive estimates from the log-normal and normal exposure models respectively. The graph shows that the  log-normal exposure model produces a distribution that is very close to that of the data, suggesting that the model adequately characterises the daily exposures. In contrast the normal exposure model  is a poor approximation to the data, having a larger variance and some posterior predictive probability below zero.   Another difference between these two models is that \cite{holloman2004} allow  only the daily exposure variance to be uncertain, with their model having the general form $\ln(\mu_{t})=\lambda_{t-l}\gamma+\mathbf{z}_{t}^{\scriptsize\mbox{T}\normalsize}\bd{\alpha},
\lambda_{t}\sim\mbox{N}(x_{t},\sigma^{2}),\sigma^{2}\sim\mbox{Uniform}(0,25)$. The posterior estimates of $\sigma^{2}$ are not informative because the Markov chains for this parameter moved quickly between the prior limits and did not converge. This lack of convergence was also observed by the authors and is presumably caused by their condensing of the simulated daily exposures into a single mean value, so that the model is trying to estimate the variation around that single value. \\

Table \ref{table_RR} also shows the  relative risks separately for the indoor and outdoor sources of pollution, which were estimated by running the pCNEM model with one of the exposure sources  turned off. From these separate simulations, the mean daily proportions attributable to indoor and outdoor sources was estimated to be 15\% and 85\% respectively. For computational reasons, the association between indoor exposures and mortality was estimated using standard regression model (ii), although this is unlikely to be of any real consequence due to  the lack of evidence of an effect of inter-day variability in this case. The table shows that the relative risk and confidence interval (and thus significance) associated with outdoor sources only is very similar to that observed with both sources combined. The risk  in relation to  indoor sources only is smaller, although this result is non-significant.\\

\begin{table}\caption{Summary of the Posterior relative risks for an increase in 10$\mu g/m^3$}
\label{table_RR}
\begin{tabular}{lllllll}
  \hline
\textbf{Data} & \textbf{Model}  & \textbf{2.5\%} & \textbf{25\%} &  \textbf{50\%} &  \textbf{75\%} & \textbf{97.5\%} \\
\hline
 Ambient & \mbox{Standard Poisson model (i)} &1.007 &1.017 &1.022 &1.027&1.037 \\ \hline
Personal & \mbox{Standard Poisson model (ii) }  &1.013 &1.039 &1.052 &1.065 &1.091 \\
Personal & \mbox{Normal exposure model (iii) } &1.023 &1.054 &1.070 &1.086 &1.118  \\
 Personal & \mbox{Log-normal exposure model (iv) } &1.013 &1.038 &1.051 &1.065 &1.090 \\ \hline
 Personal & \mbox{Outdoor sources only (iv) }  & 1.012& 1.035& 1.047& 1.059& 1.083\\
 %[OLD]
 %Personal & \mbox{Indoor sources only } &0.107& 0.280& 0.534& 0.747& 1.045& 1.951& 5.093\\
 %[NEW]Personal & \mbox{Indoor sources only } &0.140& 0.370& 0.721& 1.029& 1.466& 2.934& 8.979\\
 Personal & \mbox{Indoor sources only (ii) } &0.371& 0.729& 1.033& 1.474& 2.924\\

 \hline

\end{tabular}
\end{table}

\begin{figure}
\caption{Probabilities that the relative risk exceeds certain values. The solid line refers to the model using ambient levels, while the dotted line relates to modelled personal exposures. Panels (a) and (b) show the  results for  10$\mu$g$/m^3$ and 50$\mu$g$/m^3$ changes in PM$_{10}$ respectively.}
\scalebox{0.7}{\includegraphics{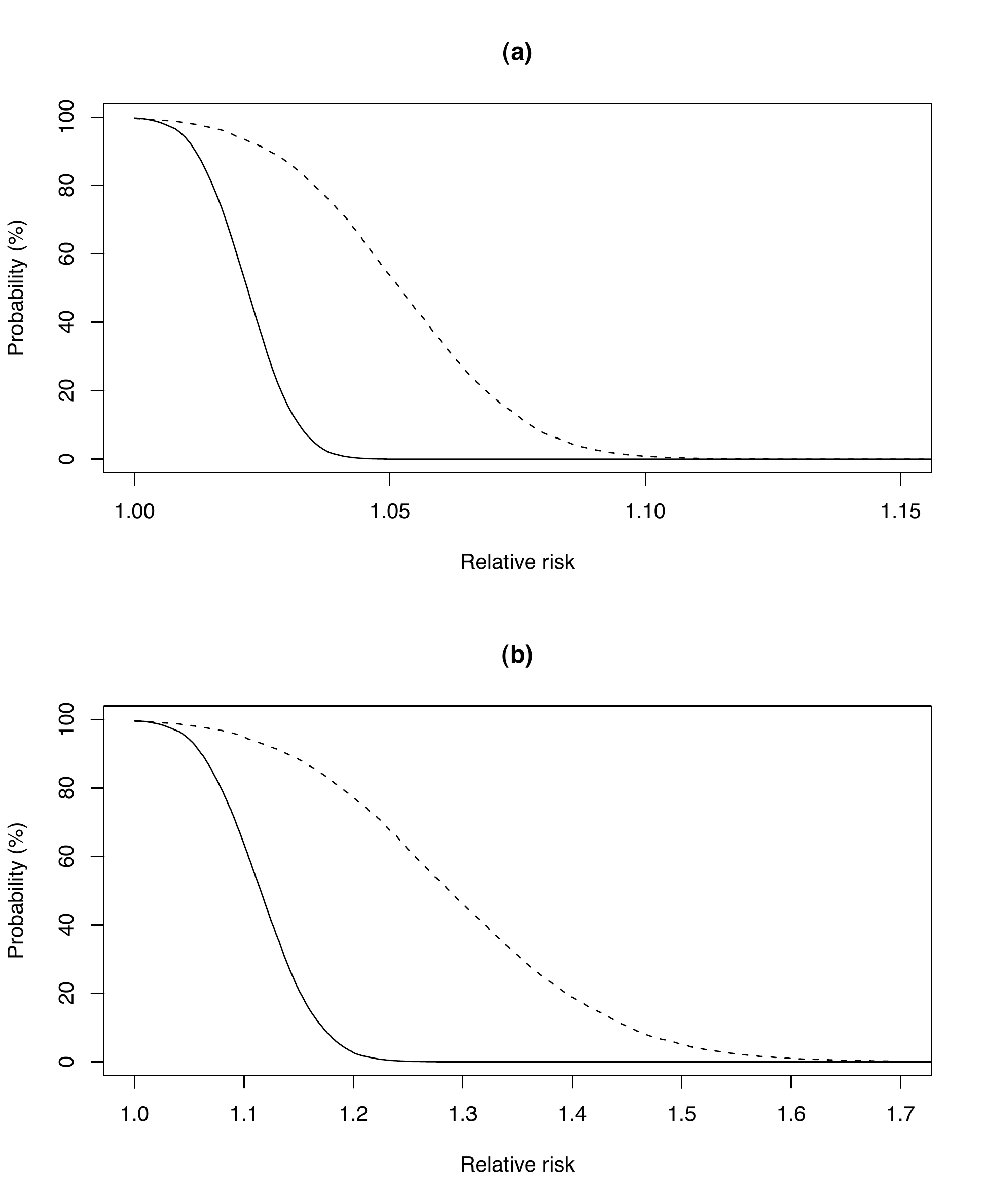}}
\label{RR_prob_graph}
\end{figure}

\subsection{The relationship between ambient concentrations and personal exposures}
In this paper the implementation of the pCNEM exposure simulator allows us to relate personal exposures to mortality, in addition to the standard use of ambient levels. Table \ref{table_RR} shows that the median relative risk from exposure to ambient levels is 1.02$\%$, less than half that obtained when personal exposures are used. The difference in the effects of personal exposures and ambient levels can be seen in Figure \ref{RR_prob_graph}, which shows $P(RR>c)$ for various values of $c$, where panel (a) relates to a relative risk for an increase in 10 $\mu g/m^3$ where as for panel (b) it is 50. The plots show clearly that $P(RR>c)$ is bigger when using personal exposures  than ambient levels, except for the case when c=1 (both probabilities are close to one)  or when $c$ is very large (both probabilities are close to 0). For example from panel (a) $P(RR>1.02) = 60.0\%$ using ambient concentrations, compared with $P(RR>1.02) = 94.3\%$ for personal exposures. This result is not surprising as the population spend a large proportion of time indoors, meaning that ambient levels are likely to be different from personal exposures leading to different relative risks with mortality, a point which is now discussed in more detail.  \\

The daily averages (means) of ambient levels and personal exposures from Greater London appear to be linearly related (see Figure \ref{graph_pcnem_data} panel (b)), with the latter being smaller by a factor of about 2.4. The same set of mortality data are used to model both pollution measures, meaning that the combined pollution-effect component of the regression model, $\lambda_{t-l}\gamma$, should remain constant regardless of the exposure size.  This relationship between the $\gamma$ regression coefficients for the ambient and personal pollution exposures holds more generally with linearly related covariates. Let $(x_{t}^{A},x_{t}^{P})$ denote ambient and personal exposures respectively, and consider the log linear models

\begin{eqnarray}
\ex{y_{t}}&=&\exp(x_{t-l}^{A}\gamma+\mathbf{z}_{t}^{\scriptsize\mbox{T}\normalsize}\bd{\alpha})\label{equation_ambient_linear}\\
\ex{y_{t}}&=&\exp(x_{t-l}^{P}\gamma^{*}+\mathbf{z}_{t}^{\scriptsize\mbox{T}\normalsize}\bd{\alpha}^{*})\label{equation_personal_linear}
\end{eqnarray}

\noindent used here where $(\gamma, \gamma^{*})$ are the parameters relating mortality to ambient levels and personal exposures respectively. Assuming the two measures of pollution are linearly related, that is $x_{t}^{P}=\theta+\phi x_{t}^{A}$, the model with personal exposures (equation \ref{equation_personal_linear}) can be re-written as

$$\ex{y_{t}}=\exp(x_{t-l}^{A}\phi\gamma^{*}+\theta\gamma^{*}+\mathbf{z}_{t}^{\scriptsize\mbox{T}\normalsize}\bd{\alpha}^{*}),$$
\noindent an alternative representation of the ambient model (equation \ref{equation_ambient_linear}). Equating the coefficients of the ambient pollution level $x_{t}^{A}$, we see that $(\gamma, \gamma^{*})$ are related as $\gamma=\phi\gamma^{*}$. Therefore if the ambient levels and personal exposures are highly correlated, then the estimated regression coefficient of the former can be determined from the latter (and vice versa) just by calculating their linear regression equation. For the Greater London data analysed here $x_{t}^{P}\approx 0.83 + 0.40 x_{t}^{A}$, meaning that $\gamma \approx 0.4\gamma^{*}$, which can be verified by comparing the posterior medians from Table \ref{table_RR}. Similar relationships are also observed by \cite{dominici2000a}, who estimate linear regressions of mean personal exposure against mean ambient levels for PM$_{10}$ from five external studies. They report estimates of $\phi$ ranging from 0.33 to 0.72, with a pooled estimate of 0.53. Recently, \cite{mcbride07} used a Bayesian hierarchical model to characterise the relationship between personal exposures and ambient concentrations of PM$_{2.5}$ for a small group of seniors in Baltimore. They also observed that using ambient concentrations would result in overestimates of personal exposure, with a mean attenuation of 0.6 (albeit with a large range). These estimates are in line with the value of 0.4 observed here, suggesting that the simulated exposures generated by the pCNEM simulator are likely to be of the correct size relative to ambient levels.

\section{Discussion}
This paper presents a two stage approach to constructing exposure response functions (ERFs), relating to the health effects of an environmental hazard monitored over time and space.  The first component uses a computer model involving ambient pollution and temperature inputs, to simulate the exposure to that hazard experienced by individuals in an urban area. The model incorporates the mechanisms that determine the level of such exposures, such as the activities of individuals in different locations which will lead to differing exposures.   The outputs from the model take the form of a set of exposures, experienced by a random sample of individuals from the population of interest for each day of the study. These daily samples can be approximated by a parametric distribution, so that the predictive exposure distribution of a randomly selected individual can be determined. The  second component incorporates these distributions into a hierarchical Bayesian framework, that jointly models the relationship between the daily exposure distributions (incorporating the within-day between individual variation) and health outcomes, whilst modelling potential confounders using splines.\\

The approach was applied to a study of the  association between particulate pollution  (PM$_{10}$) and  respiratory mortality in seniors (in London, 1997). Models using ambient concentrations and (estimated) personal exposures were compared, with the latter being represented by a single measure of pollution for each day, as well as modelling the inherent variability using both log-normal and Gaussian distributions. The use of a log-normal distribution to represent daily variability in personal exposures is more  satisfactory than the Gaussian alternative,  both in a statistical sense and in term of the physical properties of the processes that might determine concentrations. In this application the terms intended to allow for ecological bias  proved to be  negligible, meaning the health effects model was essentially log-linear and there was little difference in incorporating a (parametric) distribution for daily exposures and using a single summary measure.  As such, in this case a simpler model could have been used, although this  could not have been known \textit{a priori} and may not be true for other environmental hazards. Using the computer simulation model showed that  personal exposures to PM$_{10}$ are likely to be  significantly lower (ca. 40\%) than  measured ambient concentrations  used in regulatory standards. This implies that their relative risk (of personal exposures) is  higher than the ambient analysis would suggest (ca. 2.5 times).   The relative risk associated with  (lag two) ambient concentrations to PM$_{10}$ was RR=1.02 (1.01-1.04), with the corresponding risk associated with personal exposures being RR=1.05 (1.01-1.09). \\

This increase in observed risk is in a large part due  to the fact that the population spend a large amount of their time indoors, meaning that personal exposures (which come from indoor sources such as cooking with gas, as well as a proportion of outdoor sources determined by factors such as the air exchange rate) are likely to be lower than ambient levels (\cite{zidek2007}).  Ambient source exposures are derived from the outdoor  environment and are thus shared amongst the population, whereas non-ambient exposures come from individual environments that are not shared (\cite{sheppard2005b}). As such, careful interpretation of the meaning of the relative risk is required when comparing studies using personal exposures and ambient exposures (\cite{sheppard2005}). The traditional  time series approach relies on the assumption that  it is the (relatively) short-term temporal changes in ambient levels  that determine the relative risk coefficients  (RRCs),  and not the spatial variation in exposure to indoor sources captured by the ERF. As such, the ERF's RRCs will be (largely) determined by the ambient concentrations (Lianne Sheppard, personal communication with the third author, and also observed for the Greater London data analysed here, see Table \ref{table_RR}).  In fact, the RRCs in the CRF and ERF differ only in that the latter compensates for the lower level of predicted exposures compared with the ambient concentrations (observed for the simulated exposures generated here, the five small scale studies documented by \cite{dominici2000a} and the study of \cite{mcbride07}). For example if exposures were 50\% of ambient levels, the RRC for the ERF will have to be twice as large (since the disease effect function is roughly linear) to predict the same observed numbers of health outcomes. \\

 This  disattenuation of the RRC could be done entirely with the help of statistical models (\cite{sheppard2005}). However there will be difficulties in estimating the necessary parameters required for an entirely statistical approach, i.e.  the relationship between ambient concentrations and personal exposures for a specific sub-population, such as seniors.
The attempt to incorporate the mechanisms of how individuals are exposed rather than adopting a purely statistical approach also helps provide a more scientific basis for setting standards and analysing health effects even when in some cases the results may turn out to be similar.   The use of the computer simulation model to estimate individual exposures, and thus the ERF, therefore appears to have great potential in cases such as this, especially where the (potentially suspectable) sub-group being studied might not be expected to be well represented by using (overall) ambient levels of pollution.\\

{\bf ACKNOWLEDGEMENTS:}

The work reported here was supported by the
Natural Sciences and Engineering Research
Council of Canada. The fourth author is funded by the Bath Institute for Complex Systems (BICS), EPSRC
grant GR/S86525/01. Much of the work was done whilst the third author was a visitor at the University of Bath, funded by BICS.
\newpage

\bibliographystyle{chicago}
\bibliography{eph}
\end{document}